# Investigating IoT Middleware Platforms for Smart Application Development

Preeti Agarwal[1,a*], Mansaf Alam[1,b]

[1]Jamia Millia Islamia, New Delhi, India

`preeti.agw@gmail.com`[a], `malam2@jmi.ac.in`[b]

**Abstract.** With the growing number of Internet of Things (IoT) devices, the data generated through these devices is also increasing. By 2030, it is been predicted that the number of IoT devices will exceed the number of human beings on earth. This gives rise to the requirement of middleware platform that can manage IoT devices, intelligently store and process gigantic data generated for building smart applications such as Smart Cities, Smart Healthcare, Smart Industry, and others. At present, market is overwhelming with the number of IoT middleware platforms with specific features. This raises one of the most serious and least discussed challenge for application developer to choose suitable platform for their application development. Across the literature, very little attempt is done in classifying or comparing IoT middleware platforms for the applications. This paper categorizes IoT platforms into four categories namely: publicly traded, open source, developer friendly and end-to-end connectivity. Some of the popular middleware platforms in each category are investigated based on general IoT architecture. Comparison of IoT middleware platforms in each category, based on basic, sensing, communication and application development features is presented. This study can be useful for IoT application developers to select the most appropriate platform according to their application requirement.

## 1 Introduction

The advancement in sensor, actuator, computing and storage technologies has given rise to era of 'Internet of Things' (IoT). The emergence of smaller and cheaper interoperable wireless devices over low powered wireless medium has made the communication among these devices and human possible. These wireless devices can foster development of many smart applications in various domains such as smart home, smart traffic, smart healthcare, smart city, smart agriculture, and smart logistics etc. [26]

IoT devices sense the surroundings. If large number of IoT devices is connected, then they are going to sense large number of events, generating massive data. This data can be either structured, unstructured and semi structured. The data can be generated at constant rate or time triggered. For analysis of such kind of data, it is essential to be stored resourcefully. Different IoT vendors worldwide are coming with different middleware platforms to support application development requirement. These IoT middleware platforms mainly sit between actual deployed sensors and applications. They consist of set of functionalities such as device interaction, management, data storage, and processing required to build smart application. Broadly, IoT platforms can be divided into following four broad categories [19]: Publicly Traded, Open Source, Developer Friendly and End-to-End Connectivity platforms.

Finding most appropriate IoT middleware for an application development is the major challenge faced by the developers today. The functionalities provided by different middleware vendors are almost similar, but they differ mainly in underlying technologies. Services provided by different IoT vendors mainly include data acquisition, device management, data storage, security, and analytics. Selecting the right platform according to the detailed analysis of application requirement is one of the necessary steps in application development [7].

This paper presents some of the most popular platforms in each of the four categories. Examines their features according to the general IoT architecture. This study can help application developers in choosing the right platform according to their need. Section 2 briefly describes some studies carried related to IoT platforms., section 3 presents the general architecture of IoT platform. Section 4 presents categorization of IoT Platforms, section 5 gives comparisons of different features of IoT middleware layer wise and provide tools and technologies used in each layer. Finally, section 6 discusses criteria for choosing the right platform according to the application requirement and section 7 gives concluding remarks.

## 2   Literature Review

In this section, we will discuss studies carried out in the IoT Platform. Ray [25], studied twenty-six platforms based on ten parameters (application development, device management, system management, heterogeneity management, data management, analytics, deployment, monitoring, visualization and research).

Farahzadi et.al. [5], discussed various challenges of storage, management, aggregation of IoT data in middleware platforms. They presented a comparison of platforms on basis of overall characteristics of such as adaptability, connectivity, context management, energy efficiency, flexibility, interoperability, maintainability, platform portability, quality of service, real-time tasks, resource discovery, reusability, security and privacy, transparency, trustworthiness.

Cruz et.al. [4], classified IoT Platforms IoT Cloud Platforms based on application enablement platforms, application development platforms, device management platforms.

Singh [28], studied different hardware modules and their integration with different IoT platforms. This study focused only on hardware devices making it insufficient for application development.

Mineraud et. al. [23], evaluated open source platforms. Gap analysis is done to improve platforms and provide business opportunities.

. Robert [27], presented a comparative study of open source IoT middleware platforms. This comparison was restricted to the scalability and reliability measure. Machorro et. al. [32], carried out comparative study of platforms from industry point of view and discussed case studies with challenges. Fortino et al. [26] discussed an outline of middleware for smart objects and smart environments and compared them with specific requirements, identified in the literature.

All the above-mentioned studies have discussed parameters that give overview of different platforms but does not compare layer wise features of platforms from application development perspective. Our investigation divides the task of smart application development into layers as described in section 3 and provides layer wise features of identified platforms. Data Analytics is the heart of application development in IoT Cloud environment. Number of applications can be developed using big data analytics in cloud environment [21]. Task scheduling and load balancing is required to improve efficiency of platforms [2]. In the next section, we present the general architecture of IoT platform.

## 3    IoT Middleware Platform General Architecture

This section describes the basic elements of IoT Middleware Platform reference architecture [7]. All elements of this architecture are described in the bottom up manner as shown in fig 1.

- **Sensors:** Sensor consists of a hardware component capable of acquiring information about physical environment. This acquired information is transmitted in form of electrical signals to the connected devices. Devices connectivity can be either wired or wireless.
- **Actuator:** It is a hardware component, which receives command in form of electrical signals from connected device and performs some kid of physical action. Like sensor, it can also be connected to device either in a wired or wireless manner.
- **Device:** A Device is a hardware component consisting of processor and storage. It is connected to sensors and actuators. By the help of software, it can establish connection to IoT Integration middleware.
- **Gateway:** Sometimes gateway is required to connect device to IoT platform. Gateway is an interface that provides technologies and mechanism to interconnect between communication technologies and protocols. All devices can access gateway if they are IP enabled. Gateway is also able to store, filter and process received data before sending to cloud.

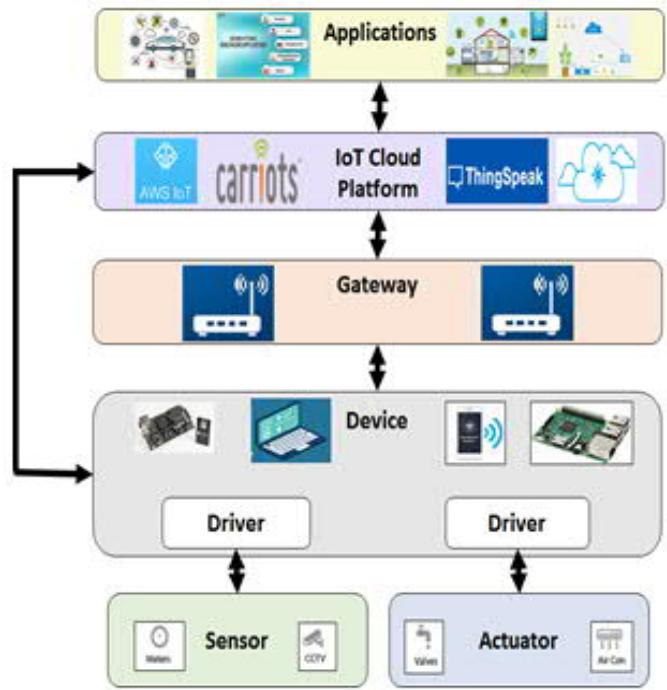

**Fig. 1.** General IoT Architecture

- **IoT Platform:** Main responsibilities of this layer are:

    a. It integrates data received from different kind of connected devices.
    b. Process the received data.
    c. Control devices
    d. Provide received data to various applications.

IoT cloud Platform can also directly communicate with device if both are using compatible technologies and protocols.

IoT Cloud Platform layer is also responsible for providing functionalities such as time series database or graphical dashboards, aggregation and utilization of data received from devices. Mostly, IoT platforms are accessed through HTTP-based REST APIs.

- **Application:** Applications are built on top of various IoT Cloud Platforms to provide services to some real-life scenario such as Smart Cities, Smart Healthcare, Smart Industry, etc.

## 4 Categorization of IoT Middleware Platforms

Various IoT Middleware platforms can be categorized into following four categories namely Public Traded IoT Cloud Platforms, Open Source IoT Cloud Platforms, Developer Friendly IoT Cloud platform, End to End connectivity IoT cloud Platform as shown in fig 2. below. This section describes various popular platforms in each of these categories:

- **Publicly Traded IoT Middleware Platforms:** This category consists of platforms developed and maintained by large public traded companies such as AWS IoT Platform [9], Microsoft Azure IoT Hub [10], IBM Watson IoT Platform [16], Google IoT Platform [11], Oracle IoT Platform [12].
- **Open source IoT Middleware Platforms**: This category consists of platforms that provide data management services under open licenses such as Kaa [17], ThingSpeak [14].
- **Developer Friendly IoT Middleware Platforms:** This category of platforms is developer friendly and can be easily integrated with Arduino, Raspberry etc. to develop users' applications. Some of the platforms belonging to this category are Carriots [15], Temboo [13].
- **End to End Connectivity IoT Middleware Platforms:** Platforms designed based on supplied hardware and required solution such as Samsara [20], Particle Cloud [18].

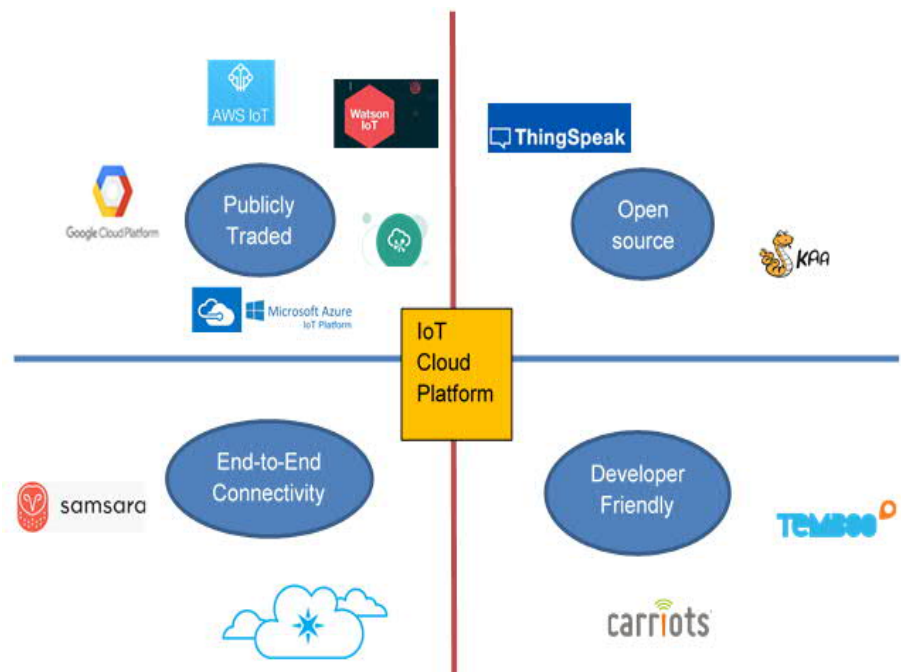

**Fig. 2.** Categorization of IoT Middleware Platform

## 5 Comparison of Features of IoT Middleware Platforms

IoT platforms in each of the above-mentioned category are compared based on basic features, sensing features, communication features, and application development features.

### 5.1 Basic Features:

The Basic features of the IoT middleware platforms include Open Source or Open API, Deployment Model of cloud, Availability, Data format supported, Programming languages supported and Pricing model. Different platforms support different features. The application developers can select a platform according to their requirements. The basic features of platforms are provided in Table 1. We have used "Y" in tables to show support of certain feature and absence of "Y" shows lack of support.

**Table 1.** Basic Features of IoT Middleware Platforms

| IoT Middleware | Open Source/Open SDK | Deployment type | Availability (24* 7) | Data Format Supported | Programming Languages Support | Pricing |
|---|---|---|---|---|---|---|
| Publicly Traded Platform | | | | | | |
| AWS IoT Platform | Open source SDK | PaaS, IaaS | Y | JSON | Java, C, NodeJS, Javascript, Python, SDK for Arduino, iOS, Android | Pay when execute your own written functions. |
| Microsoft Azure IoT Hub | Open Source API | IaaS | Y | JSON | .NET, UWP, Java, C, NodeJS, Ruby, Android, iOS | Pay according to number of devices and messages per day |
| IBM Watson IoT Platform | Open source SDK | PaaS, IaaS | Y | JSON, CSV | C#, C, Python, Java, NodeJS | Pay according to number of devices, data traffic and data storage |
| Google IoT Platform | Open API | PaaS, IaaS | Y | JSON | Go, Java, .NET, Node.js, php, Python, Ruby | Priced per MB |
| Oracle IoT Platform | Open source SDK | PaaS | Y | CSV, REST API | Java, Javascript, Android, C, iOS | Subscription based |

| Open Source Platform | | | | | | |
|---|---|---|---|---|---|---|
| Kaa | Open SDK | IaaS | - | REST API, JSON | Java, C , C++ | Free |
| ThingSpeak | Open source | PaaS | - | ThingSpeak API, JSON, XML | Matlab | Free |
| Developer Friendly Platform | | | | | | |
| Carriots | Open Source API | PaaS | Y | XML, JSON, REST API | Java | Subscription based |
| Temboo | Open source API | PaaS | - | Excel, CSV, XML, JSON | C, Java, Python, iOS, Android, Javascript | Subscription based |
| End-to-End Connectivity Platform | | | | | | |
| Samsara | Open API | - | - | JSON | - | Paid Services |
| Particle Cloud | Open Source | PaaS, Saas | - | CSV | Javascript, particle js | Free access for first 100 devices after that paid per device |

### 5.2 Sensing Features:

The Sensing features of the IoT middleware platforms include, support for multidevice, heterogeneous devices and hardware compatibility of platform for sensing environmental information. Sensing features are provided in Table 2, "Y" is used to show support of certain feature and absence of "Y" shows lack of support.

**Table 2.** Sensing Features of IoT Middleware Platform

| IoT Middleware | Multi Device Support | Heterogeneous Device Support | Hardware Compatibility |
|---|---|---|---|
| Publicly Traded Platform | | | |
| AWS IoT Platform | Y | Y | Broadcom, Marvell, Renasas, Texas Instruments, Microchip Intel |
| Microsoft Azure IoT Hub | Y | Y | Intel, Raspberry, FreeScale, Texas Instrument |
| IBM Watson IoT Platform | Y | Y | ARM mbed, Texas Instruments, Raspberry Pi, Arduino Uno |
| Google IoT Platform | Y | Y | Raspberry Pi |
| Oracle IoT Platform | Y | Y | Raspberry Pi, iMX6 sabrelite |
| Open Source Platform | | | |
| Kaa | Y | Y | Udoo, Samsung Artik, Raspberry Pi, Intel edison |
| ThingSpeak | Y | Y | Arduino, Particle photon, ESP8266 wifi, Raspberry Pi |
| Developer friendly Platform | | | |
| Carriots | Y | Y | Arduino, Raspberry Pi, Nanode, Beagle-Bone |
| Temboo | Y | Y | Texas Instrument, Arduino, Samsung artik |
| End-to-End Connectivity Platform | | | |
| Samsara | - | - | - |
| Particle Cloud | Y | Y | Electron, Photon, Raspberry Pi |

### 5.3 Communication Features:

Communication features include communication of IoT sensed data for storage and processing. Major features compared here are: Gateways, Protocols, Security mechanism. Communication Technologies used in platforms play a major role in selecting right platform for application developers. In case applications requires to connect different networks together than gateway support is must. The major communication protocols supported by IoT devices are: MQTT, CoAP, HTTP, WebSockets. Different security features supported by platforms can be encryption, authentication, authorization, auditing, and scope for user facilities. Depending on the security requirement of application the developer can choose platform. Table 3. Presents communication features, "Y" shows presence of certain feature and absence of "Y" shows absence of features.

**Table 3.** Communication Features Of Various IoT Middleware Platform

| IoT Middleware | Gateway | Protocols | Security | | | | |
|---|---|---|---|---|---|---|---|
| - | - | - | Encryption | Authentication | Authorization | Auditing | Scope for user defined policies |
| Publicly Traded Platforms | | | | | | | |
| AWS IoT Platform | - | HTTP, MQTT, Websockets | Y | Y | Y | Y | - |
| Microsoft Azure IoT Hub | Y | HTTP, AMQP, HTTP | Y | Y | Y | - | Y |
| IBM Watson IoT Platform | Y | MQTT | - | Y | Y | - | - |
| Google IoT Platform | - | MQTT, HTTP | - | Y | - | - | - |
| Oracle IoT Platform | Y | REST APIs | - | Y | Y | - | - |
| Open Source IoT Platforms | | | | | | | |
| Kaa | Y | MQTT, CoAP | Y | - | - | - | Y |
| ThingSpeak | - | MQTT | Y | - | - | - | - |
| Developer Friendly Platforms | | | | | | | |
| Carriots | Y | MQTT | - | Y | Y | - | - |
| Temboo | Y | HTTP, MQTT, CoAP | - | - | - | - | - |
| End-to- End Connectivity Platforms | | | | | | | |
| Samsara | - | HTTP | - | - | - | - | - |
| Particle Cloud | - | HTTP | Y | Y | Y | Y | - |

### 5.4 Application Development Features:

The support for various application development technologies is shown in Table 4. Application development technologies deployed on the platform is considered heart of that middleware. In the Table 4, we have discussed support for technologies such as M2M applications, real time analytics, machine learning, artificial intelligence, analytics, visualization, and event reporting. "Y" shows support for particular features and absence of "Y" means lack of support. Real time analytics refers to application of analytical algorithms on streaming data, whereas analytics refers to application of analytical algorithms on historic data.

**Table 4.** Application Development Support Features In Iot Middleware Platform

| IoT Middleware | Support for application Development | | | | | | | Technologies Used |
|---|---|---|---|---|---|---|---|---|
| | M2M application | Real Time Analytics | Machine Learning | Artificial Intelligence | Analytics | Visualization | Event and Reporting | |
| Publicly Traded Platform | | | | | | | | |
| AWS IoT Platform | - | Y | Y | Y | Y | Y | Y | AWS Lambda, Amazon Kenisis, Amazon Machine learning, Amazon Dynamo DB, Amazon CloudWatch, AWS CloudTrail, |
| Microsoft Azure IoT Hub | - | Y | Y | - | Y | Y | Y | Azure CosmosDB, Azure Tables, SQL database |
| IBM Watson IoT Platform | - | Y | Y | - | Y | Y | Y | Cloudant NOSQL DB |
| Google IoT Platform | - | Y | Y | - | Y | Y | Y | Google's BigData tool, Riptide IO, BigQuery, Firebase, PubSub |
| Oracle IoT Platform | - | Y | - | - | Y | Y | Y | NoSQL Database |
| Open Source Platform | | | | | | | | |
| Kaa | - | - | Y | - | Y | Y | Y | NoSQL, Cassandra, Hadoop and MangoDB |
| ThingSpeak | Y | Y | - | - | Y | Y | Y | Matlab, dashboard |
| Developer Friendly Platform | | | | | | | | |
| Carriots | Y | - | - | - | Y | Y | - | NoSQL Big-Database |
| Temboo | - | - | Y | - | Y | Y | Y | Microsoft Power BI, Google BigQuery |
| End-to-End Connectivity Platform ||||||||||
| Samsara | - | - | - | - | Y | Y | - | - |
| Particle Cloud | - | - | - | - | Y | - | Y | IFTTT |

## 6   Criteria for Selection of the Right IoT Middleware Platform

The most challenging task for an application developer in IoT is choosing the right platform. Each platform has certain specific features and services. The selection depends on certain criteria discussed below:

- **Availability:** Availability and stability are important parameters for application requirement. For example, Smart healthcare application requires patient medical data to be monitored continuously (24*7). Therefore, application related to patient data monitoring needs to choose a platform with 24*7 availability whereas, in case of Smart industry timings can be restricted to limited hours. As shown in table 1, most of the publicly traded platforms satisfy availability requirements.
- **Deployment type:** Open source platforms allow platforms to be managed by application developer according their need, in contrast, this facility is not provided by commercial IoT platforms such as AWS IoT, IBM Watson, etc. Professionals at the middleware provider end manage the commercial platforms. It depends on requirement of application to be developed, whether developer needs to keep flexibility of platform management on his job part or want to relax itself by levering this responsibility on vendor. Most of middleware platforms use cloud technology for storage purpose. According to the resource requirement, developer can choose among different cloud deployment models- Infrastructure as a Service (IaaS), Platform as a Service (PaaS), and Software as a Service (SaaS).
- **Pricing Model:** Different vendors adopt different pricing models. Some support pay as you execute, some pay per storage, pay according to number of connected devices, some are subscription based. For small-scale applications, some platforms provide free limited storage. Particle cloud provides free access up to 100 devices. AWS IoT charges only when function is executed on stored data. Microsoft Azure charges based on number of devices and messages. This can be suitable for applications where there is variance in number of devices and messages every time. IBM Watson charges on basis of storage required. Kaa and ThingSpeak are open sources and free, good for personalized applications. For applications with constant storage rate and devices, subscription-based platforms are good such as Carriots, Tembo.
- **Support for required Hardware:** Number of IoT boards like Arduino Yun, Photon, Raspberry Pi, etc. are available in market. Each supporting different standards and features. These boards mainly differ on basis of processor, GPU, clock speed, size, RAM, memory, support for different programming language, and price [28]. Choosing right one close to requirement is important. AWS IoT have highest number of compatible hardware devices.
- **Security Requirement:** Levels of required security vary in different applications, as banking application requires high security as compared to Smart Healthcare. Different platforms provide different level of security. Some platforms even allow application developers to implement their own security algorithms. Microsoft Azure IoT and Kaa platform provides flexibility to users to implement their own security policies. AWS IoT and Particle Cloud provide highest security features.
- **Type of communication protocol support:** Multiple communication protocols are supported by IoT devices. Some are lightweight, and some are secure [3]. Which protocol to choose depends on application requirement. CoAP is similar to HTTP but is lightweight, therefore more suitable for mobile applications. MQTT is also lightweight and supports broker concept, making it good for limited bandwidth applications [8]. CoAP is good for multicast and broadcast.
- **Storage Technologies used:** Mainly vendors use cloud as their storage. Different storage and processing technologies on top of cloud support different type of analytics. As per processing requirement of data, cloud with different storage technologies can be selected. AWS IoT supports largest number of storage technologies.
- **Type of Analytics Supported:** IoT applications usually require either Real time/Streaming data or historic data for application development [24]. Choosing right technology for applying analytics on the kind of data generated by IoT device in the application is another important factor. ThingSpeak and Kaa supports M2M applications. AWS IoT, Microsoft Azure IoT, Google IoT, Oracle IoT, ThingSpeak support real time analytics on streaming data. AWS IoT also supports artificial Intelligence.

## 7　Conclusion

In this paper, we have discussed the basic features, sensing features, communication features, and support for application development features. This information can help IoT application developers in selecting appropriate platforms according to their application need. Basic features include, support for open source/open SDK/open API, cloud deployment model used, availability, data format supported, programming languages supported and pricing model of various platforms. Sensing features tells about capability to sense information from multi devices, heterogeneous devices, compatibility with hardware. Communication features include support for gateways, different protocols supported, security features such as encryption, authorization, authentication and provision for extension of security policy. In application development fea-tures, we discussed about support of tools for application development. Different applications require different kind of tools. Lastly, we have discussed, criteria for selection of appropriate platform and the IoT middleware platforms satisfying it. This investigation can help application developers in choosing platform according to tools required in their application.

**References**

1. Al-Fuqaha, A., Guizani, M., Mohammadi, M., Aledhari, M., & Ayyash, M. (2015). Internet of things: A survey on enabling technologies, protocols, and applications. *IEEE Communications Surveys & Tutorials*, *17*(4), 2347-2376.
2. Ali, S. A., & Alam, M. (2016, December). A relative study of task scheduling algorithms in cloud computing environment. In *Contemporary Computing and Informatics (IC3I), 2016 2nd International Conference on* (pp. 105-111). IEEE.
3. Ammar, M., Russello, G., & Crispo, B. (2018). Internet of Things: A survey on the security of IoT frameworks. *Journal of Information Security and Applications*, *38*, 8-27.
4. da Cruz, M. A., Rodrigues, J. J. P., Al-Muhtadi, J., Korotaev, V. V., & de Albuquerque, V. H. C. (2018). A reference model for internet of things middleware. *IEEE Internet of Things Journal*, *5*(2), 871-883.
5. Farahzadi, A., Shams, P., Rezazadeh, J., & Farahbakhsh, R. (2018). Middleware technologies for cloud of things: a survey. *Digital Communications and Networks*, *4*(3), 176-188.
6. Fortino, G., Guerrieri, A., Russo, W., & Savaglio, C. (2014). Middlewares for smart objects and smart environments: overview and comparison. In *Internet of Things Based on Smart Objects* (pp. 1-27). Springer, Cham.
7. Guth, J., Breitenbücher, U., Falkenthal, M., Leymann, F., & Reinfurt, L. (2016, November). Comparison of IoT platform architectures: A field study based on a reference architecture. In *Cloudification of the Internet of Things (CIoT)* (pp. 1-6). IEEE.
8. Hejazi, H., Rajab, H., Cinkler, T., & Lengyel, L. (2018, January). Survey of platforms for massive iot. In *Future IoT Technologies (Future IoT), 2018 IEEE International Conference on* (pp. 1-8). IEEE.
9. Amazon Web Services, Inc. (2019). *IoT Applications & Solutions | What is the Internet of Things (IoT)? | AWS*. [online] Available at: https://aws.amazon.com/iot/ [Accessed 7 Feb. 2019].
10. Azure.microsoft.com. (2019). *IoT Hub | Microsoft Azure*. [online] Available at: https://azure.microsoft.com/en-in/services/iot-hub/ [Accessed 7 Feb. 2019].
11. Google Cloud. (2019). *Google Cloud IoT - Fully managed IoT services | Google Cloud*. [online] Available at: https://cloud.google.com/solutions/iot/ [Accessed 7 Feb. 2019].
12. Cloud.oracle.com. (2019). *Internet of Things | Oracle Cloud*. [online] Available at: https://cloud.oracle.com/iot [Accessed 7 Feb. 2019].
13. Temboo.com. (2019). *IoT :: Temboo*. [online] Available at: https://temboo.com/iot [Accessed 7 Feb. 2019].
14. Thingspeak.com. (2019). *IoT Analytics - ThingSpeak Internet of Things*. [online] Available at: https://thingspeak.com/ [Accessed 7 Feb. 2019].
15. Altairsmartworks.com. (2019). *Altair SmartWorks | Home*. [online] Available at: https://www.altairsmartworks.com/ [Accessed 7 Feb. 2019].
16. Ibm.com. (2019). *IBM Watson Internet of Things (IoT)*. [online] Available at: https://www.ibm.com/internet-of-things [Accessed 7 Feb. 2019].

17. Kaa IoT platform. (2019). *Kaa Enterprise IoT platform*. [online] Available at: https://www.kaaproject.org/ [Accessed 7 Feb. 2019].
18. Particle. (2019). *Particle Company News and Updates*. [online] Available at: https://www.particle.io/ [Accessed 7 Feb. 2019].
19. Postscapes. (2019). *IoT Cloud Platform Landscape | 2019 Vendor List*. [online] Available at: https://www.postscapes.com/internet-of-things-platforms/ [Accessed 7 Feb. 2019].
20. Samsara.com. (2019). *Samsara | Internet-Connected Sensors*. [online] Available at: https://www.samsara.com/ [Accessed 7 Feb. 2019].
21. Khan, S., Shakil, K. A., & Alam, M. (2018). Cloud-Based Big Data Analytics—A Survey of Current Research and Future Directions. In *Big Data Analytics* (pp. 595-604). Springer, Singapore.
22. Machorro-Cano, I., Alor-Hernández, G., Cruz-Ramos, N. A., Sánchez-Ramírez, C., & Segura-Ozuna, M. G. (2018). A Brief Review of IoT Platforms and Applications in Industry. In *New Perspectives on Applied Industrial Tools and Techniques* (pp. 293-324). Springer, Cham.
23. Mineraud, J., Mazhelis, O., Su, X., & Tarkoma, S. (2016). A gap analysis of Internet-of-Things platforms. *Computer Communications*, *89*, 5-16.
24. Mohammadi, M., Al-Fuqaha, A., Sorour, S., & Guizani, M. (2018). Deep Learning for IoT Big Data and Streaming Analytics: A Survey. *IEEE Communications Surveys & Tutorials*.
25. Ray, P. P. (2016). A survey of IoT cloud platforms. *Future Computing and Informatics Journal*, *1*(1-2), 35-46.
26. Razzaque, M. A., Milojevic-Jevric, M., Palade, A., & Clarke, S. (2016). Middleware for Internet of Things: A survey. *IEEE Internet of Things Journal*, *3*(1), 70-95.
27. Scott, R., & Östberg, D. (2018). A comparative study of open-source IoT middleware platforms.
28. Singh, K. J., & Kapoor, D. S. (2017). Create Your Own Internet of Things: A survey of IoT platforms. *IEEE Consumer Electronics Magazine*, *6*(2), 57-68.